\documentclass[12pt,preprint]{aastex}

\newcommand{\myemail}{zhekovs@colorado.edu}
\newcommand{\WR}{WR~48a~}
\newcommand{\WRE}{WR~48a}
\newcommand{\kms}{~km s$^{-1}$~}

\shorttitle{X-rays from \WR}
\shortauthors{Zhekov et al.}

\begin{document}


\title{
{\it XMM-Newton} Observations Reveal Very High X-ray Luminosity from
the Carbon-rich Wolf-Rayet Star \WR
}



\author{Svetozar A. Zhekov\altaffilmark{1,4},
Marc Gagn\'{e}\altaffilmark{2}, and 
Stephen L. Skinner\altaffilmark{3}
}

\altaffiltext{1}{JILA, University of Colorado, Boulder, CO
80309-0440, USA; \myemail}
\altaffiltext{2}{Department of Geology and Astronomy, West Chester
University, West Chester, PA 19383;
mgagne@wcupa.edu}
\altaffiltext{3}{CASA, University of Colorado, Boulder, CO
80309-0389, USA; stephen.skinner@colorado.edu}
\altaffiltext{4}{On leave from Space Research Institute, Sofia,
Bulgaria}


\begin{abstract}
We present {\it XMM-Newton} observations of the dusty Wolf-Rayet
star \WRE. This is the first detection of this object in X-rays. 
The {\it XMM-Newton} EPIC spectra are heavily absorbed and the
presence of numerous strong emission lines indicates a thermal origin
of the \WR X-ray emission, 
with dominant
temperature components at kT$_{cool} \approx 1$~keV and
kT$_{hot} \approx 3$~keV , the hotter component dominating the
observed flux.
No significant X-ray variability was detected on time scales 
$\leq 1$~day.
Although the distance to \WR is uncertain, if it is
physically associated with the open clusters
Danks 1 and 2 at d $\sim 4$~ kpc, then the resultant X-ray
luminosity L$_X\sim$ 10$^{35}$ ergs s$^{-1}$ makes it
the most X-ray luminous Wolf-Rayet star in the Galaxy detected so far, 
after the black-hole candidate Cyg X-3.
We assume the following  scenarios as the most likely
explanation for the X-ray properties of WR 48a: 
(1) colliding stellar winds in a wide WR$+$O binary system, 
or in a hierarchical triple system
with non-degenerate stellar components;
(2) accretion shocks from the WR 48a wind onto a close companion 
(possibly a neutron star). 
More specific information about WR48a and its wind properties
will be needed to distinguish between the above possibilities.

\end{abstract}





\section{Introduction}
As the early surveys revealed \citep{po_87}, 
Wolf-Rayet (WR) stars are sources of X-ray emission and 
massive WR$+$OB binaries are the brightest  amongst them.
Pointed X-ray observations with modern  observatories
({\it Chandra, XMM-Newton}) confirmed this and provided us with
grating spectra of WR$+$OB binaries (\citealt{sk_01};
\citealt{raa_03}; 
\citealt{schi_04}; \citealt{po_05}; \citealt{zhp_10b})
that are rich in spectral lines from different ionic species which
indicates a range  of X-ray plasma temperatures.
These findings
suggest that the enhanced X-ray emission is produced in
colliding-stellar-wind (CSW) shocks, as first proposed by
\citet{pril_76} and \citet{cherep_76}.

The available pointed {\it Chandra {\rm and} XMM-Newton} 
observations suggest 
that the X-ray properties of presumably single WR stars 
might correlate with their subtype.
Namely, the WN (nitrogen-rich) objects have X-ray spectra with 
prominent emission
lines originating from an admixture of cool (kT~$< 1$~keV) and hot
(kT~$> 2$~keV) plasma \citep{sk_10}. Only one of the four 
WO (oxygen-rich) stars in
the Galaxy, WR 142, has been observed so far and it is a weak but 
extremely hard X-ray source (\citealt{os_09};  \citealt{sokal_10}).
In contrast, the single WC (carbon-rich) stars are conspicuously 
X-ray quiet: 
all the {\it Chandra} and {\it XMM-Newton} pointed observations 
of WCs so far have resulted in non-detections 
(\citealt{os_03}; \citealt{sk_06}; \citealt{sk_09}).
The X-ray production mechanism in single WR stars
is still not well-understood, but the ubiquitous presence
of a high-temperature X-ray component in WR  spectra
is a clear indication that other processes besides embedded, 
radiative wind shocks are contributing (\citealt{sk_10}
and references therein).

Overall, binary systems, especially wide binaries, are more X-ray
luminous than comparable single WR stars. For single WR stars 
$L_X$  is typically in the range $10^{31} - 10^{33}$~ergs s$^{-1}$
\citep{sk_10}, while for the known wide WR$+$OB binaries it is
about an order of magnitude higher, with WR~140 having the largest 
known luminosity of $(1-2)\times10^{34}$~ergs s$^{-1}$ \citep{po_05}.
As a rule, if a WR star is detected with $L_{\rm X} > 10^{33}$~ergs
s$^{-1}$, it is most likely a wide CSW binary system. 

In this {\it Letter}, we report the first X-ray detection of the 
Wolf-Rayet star \WR, which establishes it as a very luminous X-ray 
source and as such, a likely multiple system.

\section{The Wolf-Rayet Star \WR}
\label{sec:thestar}
\WR was discovered in a near-infrared survey
and originally classified as a WC9 object \citep{danks_83} with a
current spectral classification of WC8 \citep{vdh_01}.
\WR is located inside the G305 star-forming region
in the Scutum Crux arm of the Galaxy. 
Within $2\arcmin$ of \WR  are two 
compact infrared clusters (Danks 1 and 2) 
and its proximity to them suggests that this WR
star likely originates from one or the other \citep{danks_84}. The 
optical extinction towards \WR is very high, A$_V = 9.2$~mag
\citep{danks_83}, with a small fraction of it ($\sim 2$~mag)
coming from circumstellar material \citep{baume_09}.

The infrared variability of \WR suggests that it is a
long-period  episodic dust-maker \citep{williams_95}, but the
dust-formation history indicates that the actual situation might be
more complex. Namely, \citet{williams_03} proposed the existence of a
short-period ($P \approx 1$~yr) binary in addition to the 
longer-period binary ($P > 23$~yr),
where the latter is thought to cause the episodic dust formation.
High-angular resolution near-infrared observations revealed a
70 mas size for the dust shell in \WR \citep{monier_07}.

The distance to \WR is not well constrained. If it is 
physically associated with
the open clusters Danks 1 and 2, 
\citet{danks_83} estimated a distance of 4 kpc, but \citet{vdh_01} 
adopted a {\it photometric} distance of 1.21 kpc based on a WC8 
spectral classification.

No information is available for the mass-loss rate and
velocity of the stellar wind in \WRE. But, indirect evidence of its
powerful wind comes from mid-infrared observations
of the region surrouding \WR \citep{clark_04}, which 
reveal a three-lobed nebula whose morphology is suggestive of a 
wind-blown structure and triggered massive star formation.

\section{Observations and Data Reduction}
\label{sec:observations}
\WR was observed with {\it XMM-Newton} on Jan 9, 2008.
As discussed below, the presence of numerous spectral lines and
our spectral analysis show that the X-ray emission is thermal and
is heavily absorbed.
The high absorption masks most emission in the RGS energy
range below 2 keV, so we were not able to obtain any 
valuable information from the grating spectra.
Thus, this study is based on 
the {\it XMM-Newton} European Photon Imaging Camera
(EPIC) data which after excluding background
flares resulted in effective exposures of 57 ksec for the pn spectrum
and 71 ksec per MOS spectrum. 
The corresponding total number of X-ray
counts in the (0.5 - 10 keV) energy range was: 75,000 (pn), 51,000
(MOS1) and 50,000 (MOS2).
We used the {\it XMM-Newton} Science Analysis System (version 10.0.0) 
to extract the source and background spectra and to 
construct the corresponding response matrices  and ancillary  
response files. 
We used standard as
well as custom models in version 11.3.2 of XSPEC
\citep{Arnaud96}
for our spectral analysis.

\begin{figure}[hp]
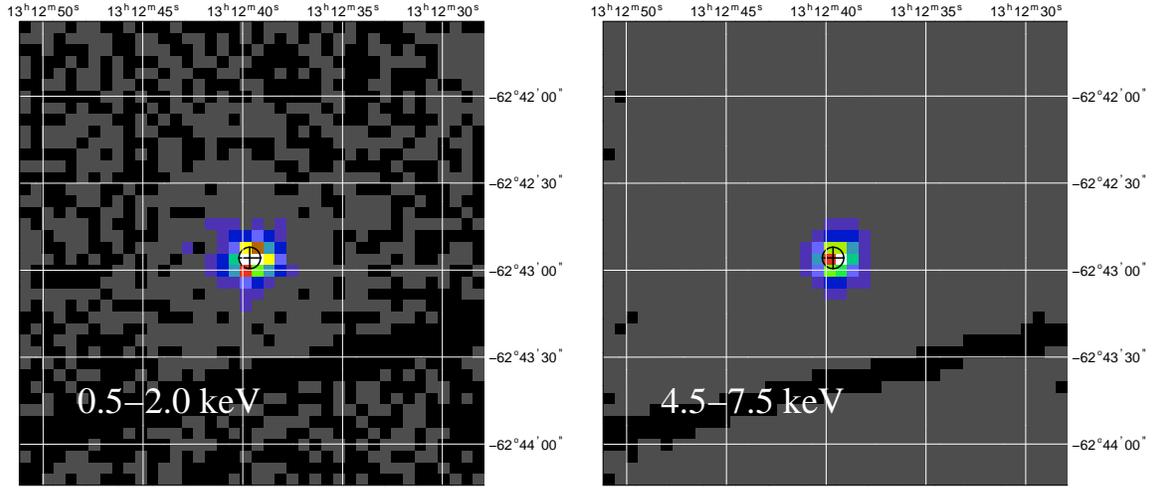

\centering\includegraphics[width=3.in,height=2.604in]{f1a.eps}
\centering\includegraphics[width=3.in,height=2.604in]{f1b.eps}
\caption{
The EPIC-pn images of \WR (linear intensity scale). 
R.A. (J2000) and decl. (J2000) are on the horizontal and vertical axes,
respectively.  The circled plus sign gives the optical position 
of \WR (SIMBAD).
}
\label{fig:images}
\end{figure}

Figure~\ref{fig:images} shows the soft and hard-band images of \WR that
illustrate an excellent correspondence between the X-ray source and
the optical coordinates of this star (J2000):
$\alpha_{XMM} = 13^h 12^m 39\fs47$,\, 
$\delta_{XMM} = -62\arcdeg 42\arcmin 56\farcs00$;
$\alpha_{SIMBAD} = 13^h 12^m 39\fs65$,\, 
$\delta_{SIMBAD} = -62\arcdeg 42\arcmin 55\farcs80$.
There is an optical
star located $\sim 9$\arcsec ~from \WR in the southwest
direction \citep{wallace_03} but we find 
no detectable X-ray emission from that source.
Thus, all the EPIC X-ray emission should be associated with \WRE
\footnote{\WR is also detected with {\it Chandra}
(\dataset[ADS/Sa.CXO\#obs/08922]{ObsId: 8922})
but pileup in the ACIS CCD is so high that no reliable spectral or
timing information for \WR  could be extracted. 
The ACIS image confirms that the \WR identification 
is correct 
(J2000: $\alpha_{Chandra} = 13^h 12^m 39\fs63,\, 
\delta_{Chandra} = -62\arcdeg 42\arcmin 55\farcs90$)
and no other sources contribute to its X-ray emission 
(Gagn\'{e} et al. 2010, in preparation).}.





\section{Global Spectral Fits}
\label{sec:global}
\begin{figure}[ht]
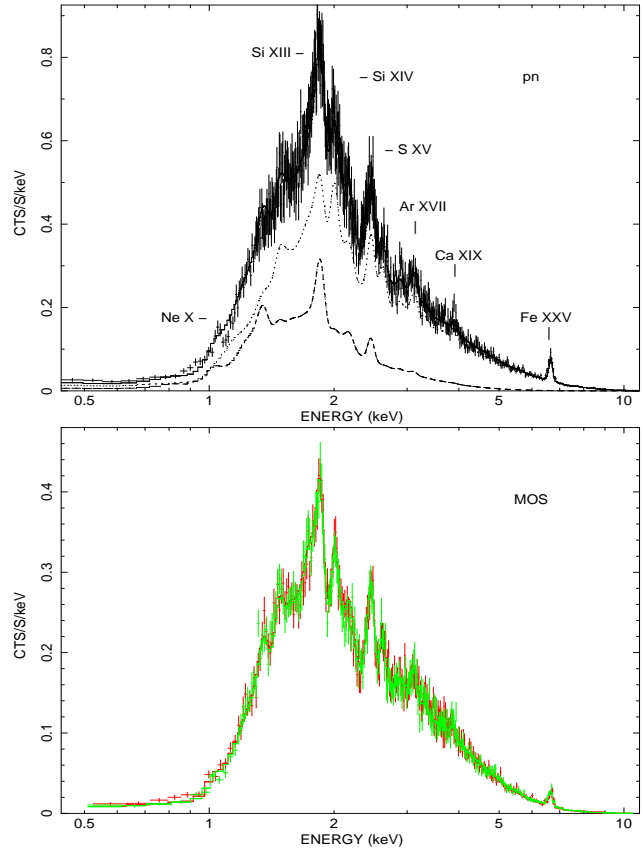

\centering\includegraphics[width=2.2in,height=3.3in,angle=-90]{f2a.eps}
\centering\includegraphics[width=2.2in,height=3.3in,angle=-90]{f2b.eps}
\caption{
{\it XMM-Newton}
background-subtracted spectra of \WR and  the two-temperature fit with
NEI shocks (Table~\ref{tab:fits}). 
The MOS1 and MOS2 spectra are drawn in red and green, respectively.
The low and high-temperature components are shown in the upper panel
with dashed and dotted lines, respectively.
}
\label{fig:spectra}
\end{figure}

A general property of the {\it XMM-Newton} spectrum of \WR
is that it is heavily absorbed and most of the X-ray photons are at
energies $> 1$~keV. We thus re-binned the pn, MOS1, and MOS2  EPIC spectra
in the 0.5 - 10 keV range to a minimum of 100 counts per bin and 
fitted them simultaneously.
We used  discrete-temperature plasma models to derive the
global properties of the X-ray emitting plasma in \WRE. We considered
two limiting cases: (i) models that assume plasma in collisional ionization 
equilibrium (CIE), and (ii)  plasma with non-equilibrium ionization (NEI).  
For consistency in the atomic data, we made  use of emission models 
{\it vapec} and {\it vpshock} in XSPEC.
Since no  specific information is available on the \WR abundances,
we adopted the following set of
abundances typical for the WC stars (by number):
H~$=0.00$, He~$=0.618$, C~$=0.248$, N~$=0.00$, O~$=0.120$, 
Ne~$=1.15\times10^{-2}$, Mg~$=1.68\times10^{-3}$,
Si~$=4.23\times10^{-4}$, S~$=9.40\times10^{-5}$,
Fe~$=2.36\times10^{-4}$
\citep{vdh_86}. 
Ar and Ca are not present in the \citet{vdh_86} abundance set, so 
assumed a fiducial value of $1.2\times10^{-5}$.
We varied the  Ne, Mg, Si, S, Ar, Ca and Fe abundances in
the spectral fits since these  elements  provide most of
the X-rays at energies $\geq 1$~keV.
Table~\ref{tab:fits} and Figure~\ref{fig:spectra} present the results
from the two-temperature model spectral fits of \WRE.

Though the quality of the two-temperature {\it vapec} model fit is 
acceptable, the NEI {\it vpshock} model provides a 
better fit to the data.
Applying the \citet{go_75} conversion
(N$_H = 2.22\times10^{21}$A$_V$~cm$^{-2}$),
we find that the X-ray absorption is consistent (within 10-15\%) 
with the optical extinction (A$_V = 9.2$~mag; \S~\ref{sec:thestar}). 
But, the X-ray data may indicate some extra absorption
($\sim 40-50$\%) if a more recent conversion is used, based on studies
of star-forming regions:
N$_H = (1.6-1.7)\times10^{21}$A$_V$~cm$^{-2}$
(\citealt{vuong_03}; \citealt{getman_05}).
The derived abundance values are 
in general consistent
with the adopted set of non-solar WC abundances. However, the
abundance values derived from X-ray analysis and especially from CCD
spectra should be considered with caution. 
Also, due to the high X-ray absorption no valuable information
could be derived for the abundances of carbon, nitrogen and oxygen.
But perhaps the most important result from the global fits is that at
least  two components with different plasma temperatures are needed
to obtain an acceptable representation of the observed spectrum,  
and the hot component requires a relatively high temperature.
The hot component (kT~$\approx 2.8 - 3.2$~keV) provides about
75-85\% of the observed flux, while the cooler component (kT~$\approx
0.9 - 1.1$~keV) provides $\sim 60$\%
of the unabsorbed X-ray emission from this WC star.

In addition to the two-component models,
we ran some more complex models: (i) a model
with a thermal CIE plasma distribution based on our custom model
which is similar to {\it c6pvmkl} in XSPEC, but uses the {\it apec}
collisional plasma for the X-ray spectrum at a given temperature; and
(ii) a model with a distribution of adiabatic NEI shocks
which made use of our custom XSPEC model  which
was successfully used in the analysis of the X-ray spectra of
SNR~1987A (e.g., \citealt{zh_09}, and the references therein).
Due to the high X-ray absorption, the X-ray spectral fits of \WR are not
sensitive to the presence of very cool plasma. Thus, we
considered only plasma temperatures above 0.5~keV for the 
emission measure distribution in \WRE.
The fit results are given in Table~\ref{tab:fits}.
Figure~\ref{fig:dem} shows that
the derived distribution of emission measure is bimodal,
with temperature peaks closely matching the components of the 
two-temperature models.

From the above results,
it seems reasonable to conclude that there is evidence for a
two-temperature distribution in the X-ray emitting region of \WRE.
But, it is worthwhile to treat such  
results with caution when based on CCD spectra
since the spectral lines are not well resolved  
and  the lines (and their ratios) are most
sensitive to the presence of a variety of plasma temperatures.
But, we believe that the bimodal distribution is in fact an
indicator of a temperature-stratified X-ray emission region in \WRE.






\section{Discussion}  
\label{sec:discussion}
\begin{figure}[ht]
\centering\includegraphics[width=3.3in,height=2.2in]{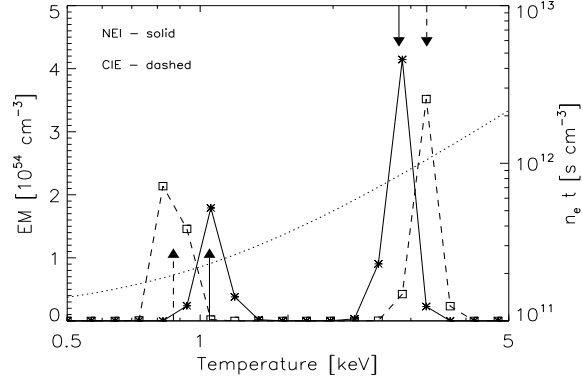}
\caption{
Emission measure ($\mbox{EM} = \int n_e n_{He} dV $) of the \WR 
distribution of thermal CEI plasma and of adiabatic NEI shocks 
for a reference distance of d~$=1$~ kpc. The
arrows indicate the plasma temperature from the two-temperature fits
(see Table~\ref{tab:fits}).
The dotted line shows
the derived ionization age in the case of NEI shocks ($n_e t$).
}
\label{fig:dem}
\end{figure}

The two most important results from the model fits to the {\it
XMM-Newton} spectra of \WR are that it is a very luminous X-ray source and 
thermal plasmas with high temperature dominate its emission 
(Table~\ref{tab:fits} and Fig.~\ref{fig:dem}). Using the
unabsorbed flux values from Table~\ref{tab:fits}, the X-ray luminosity
of \WR is L$_X (0.5 - 10~\mbox{keV}) = (0.5 - 2.1)\times10^{34} d^2_{kpc}$ 
ergs s$^{-1}$, where $d_{kpc}$ is the distance in units of kpc.
The upper limit is from the NEI shock model which  gives a better
quality of  fit. 
The distance to \WR is not tightly constrained
and a range of values is available in the literature: 
$ d_{kpc} = 1.21 - 4$ (\S~\ref{sec:thestar}). But,  
based on its proximity to the open clusters Danks 1 and 2 and the 
similar interstellar extinction (e.g., \citealt{danks_83}; 
\citealt{clark_04}; \citealt{baume_09}), the distance of 4 kpc seems 
more realistic, which puts the X-ray luminosity in the range 
L$_X = (0.8 - 3.4)\times10^{35}$ ergs s$^{-1}$.
In this case and excluding the most X-ray luminous WR star, Cyg X-3,
which is a WR binary with a compact companion (a neutron star or a
black hole), \WR is  the most luminous WR star in the Galaxy
among those observed so far.
For example, the X-ray luminosity of the brightest CSW binary, WR 140,
is: $(0.5 - 3)\times10^{34}$ ergs s$^{-1}$ (\citealt{zhsk_00};
\citealt{po_05}).
Adopting the bolometric luminosity from \citet{clark_04}, we note the
very high value of $\lg L_X/L_{bol} = [-4.3, -3.7]$ for \WRE.
All this raises the interesting question about the X-ray 
production mechanism in \WRE.

{\it Colliding Stellar Winds.}
As already mentioned, \WR is an episodic dust-maker \citep{williams_95} which
suggests that CSWs in a wide binary system with an eccentric orbit might 
provide most of its X-ray emission, as is the case for the prototype
episodic dust-maker amongst WR stars, WR 140 \citep{williams_90}. 
Unfortunately, the lack of information about the
stellar wind parameters of \WR did not allow us to make a comparison
between the theoretical predictions based on hydrodynamic CSW modeling 
and observations. On the other hand, some simple (e.g. {\it
qualitative}) considerations are possible. The temperature of the hot
plasma component deduced from the global spectral fits (\S
\ref{sec:global}) can provide an estimate of the stellar wind
velocity (see \S 5.2 in \citealt{zh_07} for discussion of CSW models 
versus discrete-temperature models ). For the case of typical WC
abundances, the postshock plasma temperature is kT~$ = 3.09
~V_{1000}^2$~keV, where $V_{1000}$ is the shock velocity in units of
1000\kms. From Table~\ref{tab:fits} and Fig.~\ref{fig:dem}, we
see that the stellar wind of the WC star in \WR must have a velocity
V$_{wind} \geq 1000$\kms.
Such high velocities would be expected
for WC8-9 stars since they have typical average wind speeds
1400\kms \citep{prinja_90}, 1300\kms \citep{eenens_94}.
We note that the derived values for ionization age of the shocks in the
two-temperature model (Table~\ref{tab:fits}) and in the model with a
distribution of NEI shocks (Fig.~\ref{fig:dem}) are also qualitatively
consistent with the CSW picture. Namely, the higher density
and higher temperature plasma is located near the axis of symmetry
(the line-of-centers between the two stars)
while the less dense and cooler plasma is found downstream from that
axis. Thus, it is natural in the CSW picture that the higher temperature 
plasma will have a larger ionization age.
We can use the scaling law for the CSW X-ray luminosity with the
mass-loss rate ($\dot{M}$), wind velocity ($v$) and binary separation 
($D$): $L_X \propto \dot{M}^2 v^{-3} D^{-1}$ (\citealt{luo_90};
\citealt{mzh_93}) for a comparison between \WR and the classical CSW
binary WR 140. Adopting the WR 140 binary parameters
(\citealt{williams_90}; \citealt{po_05}), a wind velocity 
of $1300-1400$ \kms in \WR and  assuming that its mass loss is equal
to that of WR 140,
we see that the binary separation in \WR could be similar to that 
in WR 140 even when the \WR X-ray luminosity is about an order of
magnitude larger than that of WR 140 (see above).
Therefore, the CSW picture in a wide binary system is at least
qualitatively consistent with the observational data for \WRE.

{\it Magnetically Confined Wind Shocks (MCWS).}
This mechanism is capable of producing hard X-ray emission. It was
proposed to explain the high plasma temperature found in  X-ray
emission from young massive stars and requires the presence of a relatively
strong  magnetic field (\citealt{sch_03}, \citealt{gagne_05} and 
references therein). Then, could it be that the WC star in \WR is 
a magnetized object or  has a close magnetized companion?
We note that the X-ray luminosity of the prototype MCWS
object, $\theta^1$~ Ori C, is 
L$_X (0.5 - 10~\mbox{keV}) \approx 10^{33}$ ergs s$^{-1}$
\citep{gagne_05} which is one to two orders of magnitude below that
of \WRE. Thus, we can likely rule out the case of a close
magnetized companion. On the other hand, the MCWS mechanism does not
seem promising if adopted directly to the WC star in \WRE. First,
in order to be efficient this mechanism requires a relatively strong 
global magnetic field to confine the massive WR wind and no such fields
have been reported at present for these stars. Second, because of the
expected decay of the magnetic field strength with the age of a
massive star, the MCWS mechanism is associated only with young massive
stars (age $\leq 1$~Myr; \citealt{sch_03}). Thus, the  likely
association of \WR with the open clusters Danks 1, 2 and
the recent estimate of $\sim 5$~Myr for their age \citep{baume_09}
present serious difficulties for the MCWS model in the case of \WRE.

{\it Wind Accretion Shocks.}
The presence of a close degenerate companion (e.g. a neutron star) seems very
intriguing and accretion onto such an object can in general provide
high X-ray luminosities (e.g., \citealt{do_73}). Details depend on the
actual binary parameters and wind properties of the main stellar
component which are yet unknown for \WRE. 
But, this case may not be unlikely if \WR is indeed 
a member of the 5-Myr-old open clusters Danks 1, 2
and if it once had a  more massive close companion which 
has already evolved and exploded as a supernova.

We have listed above some physical mechanisms that might be able to
explain the observed X-ray properties of \WRE. One could
continue this list by considering other mechanisms or even more
complex combinations of those mentioned above. But, all such 
mechanisms and the corresponding physical picture remain quite
speculative because of the scarcity of 
detailed observational data for \WR.
Thus, we will end our discussion by simply mentioning the cases
that look least speculative to us and briefly discuss what
observational information may help distinguish between them.

For the moment, the following cases are 
the most likely explanation for the X-ray properties of \WRE:
(i) CSWs in a wide binary system; or
(ii) accretion wind shocks in a close binary: WR$+$compact
companion (a neutron star).
Variability is a key parameter for distinguishing between 
these cases. Namely, if  long-term (months, years) X-ray variability 
is established from future observations, then case (i) would be strongly 
favored.  Furthermore, case (i) would be strengthened
if  variable non-thermal radio (NTR) emission from  \WR is eventually 
discovered and even more so if the X-ray emission is found to modulate
at a similar period. In general, NTR WRs are commonly associated with wide
binary systems (\citealt{do_00}).
Alternatively, if  short-term (a few days) variability is 
established  or a sudden change of the X-ray luminosity is detected, 
case (ii) would be favored. 
The {\it XMM-Newton} data show no significant variability on 
timescales $\le 1$~day; the Kolmogorov-Smirnov test rules out variability 
at the 95\% confidence level.
On the other hand, based on the infrared variability of
\WRE, \citet{williams_03} proposed that it is a triple
stellar system. If so, we may expect some very long-period
variability due to the X-rays from CSWs in a wide binary with eccentric
orbit and some with a much shorter period that could  result from the
stellar wind shocking onto a non-degenerate companion. 
In this case, WR 48a would be very similar to the CSW
binary WR 147 which was recently resolved into a double
X-ray source with {\em Chandra}, of which one component
is variable and possibly an unresolved binary (\citealt{zhp_10a},b).
However, WR 48a has a much higher X-ray luminosity.
Finally, we hope that the unusually high L$_{X}$ of \WR will
motivate deeper follow-up observations across the entire 
spectrum (radio, infrared, optical, UV, X-rays). Such observations
would  provide us with valuable information about the star 
and its wind properties that could be very helpful for constraining the 
physical picture in this remarkable but understudied  Wolf-Rayet 
system.




\acknowledgments
This work was supported by NASA through {\it Chandra} Award GO8-9014X 
and NASA Award NNX08AO69G to the West Chester University, West Chester,
Pennsylvania.
SAZ acknowledges financial support from Bulgarian National Science
Fund grant DO-02-85.
The authors thank an annonymous referee for his/her comments and
suggestions.



{\it Facilities:} \facility{{\it XMM-Newton} (EPIC)}.

\clearpage



%

\clearpage

\clearpage

\clearpage

\begin{deluxetable}{lllll}
\tablecaption{Global Spectral Model Results 
\label{tab:fits}}
\tablewidth{0pt}
\tablehead{
\colhead{Parameter} & \colhead{2T vapec}  & \colhead{2T vpshock} &
\colhead{CIE Plasma}  & \colhead{NEI Shocks}
}
\startdata
$\chi^2$/dof  & 1382/1066 & 1169/1064 & 1403/1064 &  1173/1062 \\ 
N$_{H}$ (10$^{22}$ cm$^{-2}$)  & 
          2.27$^{+0.03}_{-0.02}$ & 2.30$^{+0.05}_{-0.04}$ & 
          2.29$^{+0.03}_{-0.03}$ & 2.29$^{+0.04}_{-0.04}$ \\ 
kT$_1$ (keV) & 0.87$^{+0.01}_{-0.01}$ & 1.05$^{+0.09}_{-0.07}$ & & \\ 
kT$_2$ (keV) & 3.26$^{+0.06}_{-0.07}$ & 2.82$^{+0.07}_{-0.06}$ & & \\ 
EM$_1$ ($10^{54}$~cm$^{-3}$) &  3.58$^{+0.10}_{-0.07}$ &  
                                2.42$^{+0.08}_{-0.16}$ & & \\
EM$_2$ ($10^{54}$~cm$^{-3}$) &  4.21$^{+0.06}_{-0.04}$  & 
                                5.34$^{+0.04}_{-0.04}$ & & \\
$\tau_1$ ($10^{11}$ cm$^{-3}$ s)  &   &   2.42$^{+0.56}_{-0.42}$ & & \\ 
$\tau_2$ ($10^{11}$ cm$^{-3}$ s)  &   &   8.09$^{+1.39}_{-1.02}$ & & \\ 
Ne  & 0.46$^{+0.10}_{-0.09}$ & 0.11$^{+0.03}_{-0.01}$  
    & 0.53$^{+0.16}_{-0.12}$  & 0.11$^{+0.03}_{-0.03}$  \\ 
Mg  & 0.23$^{+0.04}_{-0.03}$ & 0.13$^{+0.02}_{-0.02}$  
    & 0.25$^{+0.04}_{-0.04}$  & 0.13$^{+0.02}_{-0.02}$  \\ 
Si  & 1.35$^{+0.06}_{-0.06}$ & 0.64$^{+0.06}_{-0.05}$  
    & 1.38$^{+0.07}_{-0.07}$  & 0.65$^{+0.06}_{-0.06}$  \\ 
S   & 4.90$^{+0.23}_{-0.20}$ & 1.78$^{+0.15}_{-0.12}$  
    & 4.88$^{+0.20}_{-0.20}$  & 1.81$^{+0.12}_{-0.12}$  \\ 
Ar  & 9.72$^{+0.94}_{-0.92}$ & 2.78$^{+0.19}_{-0.36}$  
    & 9.58$^{+1.03}_{-0.94}$  & 2.82$^{+0.38}_{-0.34}$  \\ 
Ca  & 5.19$^{+0.65}_{-0.62}$ & 1.97$^{+0.34}_{-0.33}$  
    & 5.22$^{+0.69}_{-0.70}$  & 2.00$^{+0.33}_{-0.41}$  \\ 
Fe  & 1.36$^{+0.06}_{-0.07}$ & 1.31$^{+0.07}_{-0.07}$  
    & 1.37$^{+0.07}_{-0.07}$  & 1.30$^{+0.07}_{-0.07}$  \\ 
F$_{X}$ ($10^{-11}$ ergs cm$^{-2}$ s$^{-1}$)  & 
           1.06 (4.21) & 1.05 (17.7) & 
           1.06 (4.30) & 1.05 (17.8) \\ 
F$_{X,hot}$ ($10^{-11}$ ergs cm$^{-2}$ s$^{-1}$)  & 
           0.82 (1.79) & 0.89 (6.98) & 
           0.81 (1.79) & 0.89 (6.81) \\ 
\enddata
\tablecomments{
Results from  simultaneous fits to the EPIC 
spectra of \WRE.
Tabulated quantities are the neutral hydrogen absorption column
density (N$_{H}$), plasma temperature (kT), 
emission measure ($\mbox{EM} = \int n_e n_{He} dV $; see also
Fig.~\ref{fig:dem})
for a reference distance of d~$=1$~ kpc, 
shock ionization age
($\tau = n_e t$), the absorbed X-ray flux (F$_X$) in the 
0.5 - 10 keV range followed in parentheses by the unabsorbed value
(F$_{X,hot}$~ denotes the higher-temperature component, kT$_2$; or
kT$ \geq 2$~keV for CIE Plasma and NEI Shocks).
The derived abundances are with
respect to the typical WC abundances \citep{vdh_86}.
Errors are the $1\sigma$ values from the fits.
}

\end{deluxetable}




\end{document}